\documentstyle[11pt]{article}
\begin{document}
%\begin{titlepage}
\vspace{0.5cm} \centerline{\Large \bf Gapless formation in the $K^0$
condensed color-flavor locked quark} \centerline{\Large\bf   matter
: a model-independent treatment} \vspace{1cm} \centerline{Xiao-Bing
Zhang $^{1,2}$ and J. I. Kapusta $^{2}$} \vspace{0.1cm}
\centerline{\small $^{1}$ Department of Physics, Nankai University,
Tianjin 300071, China} \centerline{\small $^{2}$ School of Physics
and Astronomy, University of Minnesota, Minneapolis, Minnesota
55455} \vspace{8pt}

\vspace{0.5cm}

\begin{minipage}{13cm}
%\centerlineWhenelectron chemical potential is nonzero
{\rm \noindent The electric/color neutral solution and the critical
conditions for gapless formation are investigated in the $K^0$
condensed color-flavor locked matter. We point out that there exist
no longer gapless modes for down-strange quark pairing while the
gapless phenomenon for up-strange one is dominated in the $K^0$
condensed environment. In a model-independent way, the phase
transition to the resulting gapless phase is found to be of
first-order. The novel phase structure implies that the
chromomagnetic instability happens in the previous-predicted gapless
phase might be removed at least partly.}

\vspace{0.5cm} {\bf PACS number(s): 25.75.Nq, 12.39.Fe, 12.38.-t}

\end{minipage}

\baselineskip 18pt

\vspace{0.2cm} \noindent {\bf I. INTRODUCTION} \vspace{0.2cm}

In recent years, the studies of dense matter become a "hot" topic
since quark color superconducting phases were proposed for
high-density QCD. For three flavors, the original color and flavor
$SU(3)_{color}\times SU(3)_{L} \times SU(3)_{R}$ symmetries of QCD
are broken down to the diagonal subgroup $SU(3)_{color+L+R}$ at very
high baryon density \cite{alf}. The quark matter with such a
particular symmetry pattern is called color-flavor locked ( CFL )
matter which is widely believed to be the densest phase of strongly
interacting matter \cite{alf03}. In the situation where the strange
quark mass $m_s$ is large and/or the quark chemical potential $\mu$
is small, the so-called gapless color-flavor locked phase ( gCFL )
has been predicted \cite{liu,alf04}. Similar as the gapless phase of
two-flavor color superconductor \cite{huang}, gCFL is triggered by
mismatches between the chemical potentials for paired quasiquarks.
There, the mismatch in the flavor ( say $d$ and $s$ ) chemical
potentials is characterized by $m_s^2/2\mu$ directly while that in
the color ( say $b$ and $g$ ) chemical potentials is related with
$m_s^2/2\mu$ via the electric/color neutrality in CFL matter
\cite{alf04}. As long as $m_s^2/2\mu$ is large enough, therefore,
there exist the unpaired ( gapless ) modes for $bd$-$gs$ pairing and
the resulting gCFL phase becomes energetically favorable. Recently
the Meissner masses for some gluons in this phase were found to be
imaginary so that gCFL is actually unstable in the chromomagnetic
sense \cite{cas05,fuku,alfjp05}. The presence of chromomagnetic
instability is a serious problem so that the gapless phase of CFL
matter especially its  needs to be further examined.
%As a candidate for the second-densest phase of matter \cite{alf05},

On the other hand, another kind of less-symmetric phase of CFL
matter was predicted in the situation of $m_s \neq 0$ also. At the
leading order, it is practical to regard the effect of $m_s^2/2\mu$
as an effective chemical potential associated with strangeness,
namely
\begin{eqnarray} \mu_S = \frac{m_s^2}{2\mu}. \label{muk}
\end{eqnarray}
As one of the pseudo Goldstone bosons related to the CFL symmetry
pattern, the neutral kaon mode $K^0$ obtains its chemical potential
$\mu_{K^0}=\mu_S $ due to the chemical equilibrium. When $\mu_S$
exceeds the mass of kaon-mode, $K^0$ becomes condensed in CFL matter
\cite{sch,kr}. If ignoring the instanton contribution to the kaon
mass, it was found that the $K^0$ condensed phase of CFL matter (
CFL$K^0$ ) becomes possible as long as $m_s$ is larger than
%(\frac{2\sqrt{3}\mu}{\pi f_\pi})^{2\over 3} m_q^{1\over 3}
%\Delta_0^{2\over 3}\approx \geq
the critical value $\sim m_q^{1/3} \Delta_0^{2/3}$ \cite{sch}.
Because the light quark mass $m_q$ is far smaller than $m_s$ and the
CFL gap $\Delta_0$ has order of tens MeV, CFL$K^0$ is usually
favored over CFL.
%It is necessary and might be important to investigate the two kinds
%of phases, namely CFL$K^0$ and gCFL, at the same time.
%As a whole the above physical picture is correct, but its
%details are warrant further investigation.
Starting at the ideal situation of $m_s=0$ and raising $m_s^2/2\mu$
gradually, one expects that CFL matter to be disrupted, at first by
the presence of $K^0$ condensation and then by the gapless
formation. In the CFL matter with $K^0$ condensation, the gapless
formation has been reexamined by using an effective theory
\cite{kry} and the NJL model \cite{forbes,buba}. It was found the
gapless formation is delayed with respect to that in the
conventional CFL matter. More recently, it has been suggested that
the chromomagnetic instability in the gapless phase can be resolved
by the formation of the so-called kaon supercurrent state
\cite{sch06}.

Motivated by these results, we would like to investigate the gapless
formation in the CFL$K^0$ environment through a model-independent
approach. Firstly, we consider the electric/color neutral solution
of CFL$K^0$ in the presence of electron chemical potential and point
out that the above-mentioned delay is a direct consequence of the
deviation of the CFL$K^0$ neutral solution from the CFL one. Then,
it is found that there exist no longer the gapless modes for $bd-gs$
pairing while the gapless phenomenon for $bu-rs$ pairing becomes
dominated in the $K^0$ condensed environment. Based on this feature,
the resulting gapless phase ( termed as gCFL$K^0$ below ) including
its electric/color neutrality and gap variation are studied. As a
consequence, a first-order phase transition from CFL$K^0$ to
gCFL$K^0$ is obtained. Finally, the stability of the gCFL$K^0$ phase
is examined qualitatively and it is argued that the gCFL instability
might be removed at least partly.
%%%%%%
% and thus might be a morereasonable candidate for the second-densest strongly interactingmatter.
%Since is deviated from the CFL solution, the changes in the
%mismatches of chemical potentials and thus the critical conditions
%for gapless formation are natural. Along this line,

\vspace{0.2cm}\noindent {\bf II. ELECTRIC/COLOR NEUTRALITY IN THE
$K^0$ CONDENSED ENVIRONMENT }\vspace{0.2cm}

%\vspace{0.1cm}\noindent {\bf A.  Electric/color neutral solutions :
%from CFL to CFL$K^0$}  \vspace{0.1cm}

The stable, bulk matter must be electrically neutral and must be a
color singlet, i. e. its free energy $\Omega$ satisfies
\begin{eqnarray}
\frac{\partial \Omega}{\partial \mu_e}=\frac{\partial
\Omega}{\partial \mu_3}=\frac{\partial \Omega}{\partial \mu_8}=0,
\label{neu}
\end{eqnarray}
where $\mu_e$, $\mu_3$ and $\mu_8$ are the chemical potentials
coupled to the negative electric charge, the color generators $T_3$
and $2 T_8/\sqrt{3}$ respectively. For the CFL matter, the
electric/color neutral solution has been obtained from the
derivative of $\Omega_{CFL}(\mu_e,\mu_3,\mu_8)$ in a
model-independent way and it can be expressed by \cite{alfj02}
\begin{eqnarray} \mu_3=\mu_e, \,\,\,\,\,\,
\mu_8=\frac{\mu_e}{2}-\frac{m_s^2}{2\mu}\equiv
\frac{\mu_e}{2}-\mu_S, \label{ncfl}
\end{eqnarray}
to order $m_s^2$. Another important feature involving CFL is that
its free energy is independent of
\begin{eqnarray} \mu_{\tilde{Q}}=-\frac{4}{9}(\mu_e+\mu_3+\frac{\mu_8}{2}),
\label{Q}
\end{eqnarray}
since all the CFL pairings do not break the rotation
${\tilde{Q}}=Q-T_3-\frac{1}{\sqrt{3}}T_8$. In this sense, CFL is not
merely an electric insulator \cite{rw01} but also a ${\tilde{Q}}$
insulator \cite{alfj02}.

For the matter with $K^0$ condensation, its free energy
$\Omega_{CFLK^0}$ makes up of the CFL free-energy contribution and
the contribution from $K^0$ condensation. Since the condensation
does not bring in any additional electric and color charges, the
electric/color neutral solution in CFL$K^0$ may be obtained from the
derivative of the former contribution. Superficially, it seems
obvious that the CFL$K^0$ solution has the same form as
Eq.(\ref{ncfl}).
%It seems simple to yield $\mu_3$ and $\mu_8$ from ${\partial
%\Omega_{CFLK^0}}/{\partial \mu_i}={\partial \Omega_{CFL}}/{\partial
%\mu_i}=0$ ( $i=3, 8$ ) since Eq.(\ref{pcon}) .
However, this is not correct yet. In the presence of $K^0$
condensation, the ground state of Goldstone bosons takes non trivial
values so that the corresponding vacuum should be changed with
respect to the conventional CFL vacuum. By treating the gauge fields
as dynamical degrees of freedom, Kryjevski has pointed out that
there exist nonzero vacuum expectation values for the gluon and
photon fields in the CFL$K^0$ matter \cite{kry03}. It was found that
the electric charge neutrality leads to that the time components of
two color-diagonal gluon fields are
\begin{eqnarray} \langle
A^0_3\rangle=\frac{m_s^2}{4g\mu},\,\,\,\,\,\,\,\langle A^0_8 \rangle
=\frac{m_s^2}{4\sqrt{3}g\mu},\label{a38}
\end{eqnarray}
where $g$ is the QCD coupling coefficient. Due to the non-vanishing
vacuum, the CFL free-energy contribution in the $K^0$ condensed
environment is not equal to $\Omega_{CFL}(\mu_e,\mu_3,\mu_8)$
defined in CFL matter. In this case, the electric/color neutral
solution in CFL$K^0$ obtained from Eq.(\ref{neu})  is expected to be
different from the well-known solution Eq.(\ref{ncfl}).

In order to yield the CFL$K^0$ neutral solution, we need to take
such a non-vanishing vacuum as Eq.(\ref{a38}) into account
self-consistently. Note that introducing Eq.(\ref{a38}) means that
there exist terms like
\begin{eqnarray}
g\langle A^0_3\rangle \psi^+ T_3 \psi=\frac{\mu_S}{2}\psi^+ T_3
\psi, \nonumber \\
g\langle A^0_8\rangle \psi^+ T_8 \psi=\frac{\mu_S}{4}\psi^+
\frac{2}{\sqrt{3}} T_8 \psi, \label{a382}
\end{eqnarray}
in the effective Lagrangian describing quasiquarks. By combining
Eq.(\ref{a382}) with the regular terms involving $\mu_3$ and
$\mu_8$, it is convenient to define the fictional chemical
potentials
\begin{eqnarray} \mu'_3=\mu_3+\frac{\mu_S}{2},\,\,\,\,\,\,\,\,
\mu'_8=\mu_8+\frac{\mu_S}{4},\label{nu'}
\end{eqnarray}
which are associated with the new gluon fields with vanishing
vacuum.  With the help of Eq.(\ref{nu'}), the CFL free-energy
contribution in the $K^0$ condensed environment behaves as the
function of $\mu'_3$, $\mu'_8$ and $\mu_e$ formally. In analogy with
the treatment in Ref.\cite{alfj02}, it is easily found that
\begin{eqnarray}
\frac{\partial \Omega_{CFLK^0}(\mu_e,\mu'_3,\mu'_8)}{\partial
\mu'_3}=\frac{\mu^2}{4\pi^2}(\mu_e-\mu'_3),
\nonumber \\
\frac{\partial \Omega_{CFLK^0}(\mu_e,\mu'_3,\mu'_8)}{\partial
\mu'_8}=\frac{\mu^2}{6\pi^2} (\mu_e-2\mu'_8),\label{neueq}
\end{eqnarray}
at the leading order. By requiring Eq.(\ref{neueq}) to be zero (
color neutrality ), the factual color chemical potentials become
\begin{eqnarray} {\mu_3}=\mu_e-\frac{\mu_S}{2},
\,\,\,\,\,\, {\mu_8}=\frac{\mu_e}{2}-\frac{\mu_S}{4},\label{ncflk}
\end{eqnarray}
where Eq.(\ref{nu'}) has been used. In addition, Eq.(\ref{ncflk})
makes the derivative
\begin{eqnarray}
\frac{\partial \Omega_{CFLK^0}(\mu_e,\mu_3,\mu_8)}{\partial
\mu_e}=\frac{\mu^2}{12\pi^2}(2\mu_S+3\mu_3+2\mu_8-4\mu_e),\label{neueq'}
\end{eqnarray}
to be zero automatically since the electric neutrality has been
considered in Eq.(\ref{a38}). As expected, the CFL$K^0$ solution
Eq.(\ref{ncflk}) is deviated from the CFL solution. Interestingly,
the $\mu_e$-dependence in the two solutions remain to have the same
forms, which manifests that $K^0$ condensation is independent of
electric charge and CFL$K^0$ is still an insulator. Eq.(\ref{ncflk})
can be understood as the extrapolation of the solution
${\mu_3}=-\mu_S/2$ and ${\mu_8}=-\mu_S/4$ obtained in
Ref.\cite{forbes} to the case of $\mu_e \neq 0$.

%\vspace{0.1cm}\noindent {\bf B. Mismatches in the chemical
%potentials :  CFL \emph{vs.} CFL$K^0$ } \vspace{0.1cm}

Ignoring the diquark condensates which are symmetric in color, we
consider the pairing ansatz in the form of
\begin{eqnarray} {\langle \psi^\alpha_i C\gamma_5
\psi^\beta_j\rangle} \sim \Delta_1 \epsilon^{\alpha\beta
1}\epsilon_{ij1} + \Delta_2 \epsilon^{\alpha\beta 2}\epsilon_{ij2} +
\Delta_3 \epsilon^{\alpha\beta 3}\epsilon_{ij3} ,\label{ansatz}
\end{eqnarray}
where $(i,j)$ and $(\alpha,\beta)$ denote the flavor indices
$(u,d,s)$ and the color indices $(r,g,b)$ respectively and the gap
parameters $\Delta_1$, $\Delta_2$ and $\Delta_3$  are approximately
equal to $\Delta_0$ in the absence of gapless phenomena. Based on
Eq.(\ref{ansatz}), the $bd-gs$, $bu-rs$ and $rd-gu$ pairings are
described by the gaps $\Delta_1$, $\Delta_2$ and $\Delta_3$
respectively ( termed as I, II and III channels for short ), while
the $ru$, $gd$ and $bs$ quarks pair among each other in a way
involving all three gaps. For the three channels, the mismatches
between the chemical potentials for paired quarks are given by
\begin{eqnarray} \delta \mu_I &=& \frac{\mu_{bd}-\mu_{gs}}{2}=
\frac{\mu_3}{4}-\frac{\mu_8}{2}+\frac{\mu_S}{2},\nonumber \\
\delta \mu_{II}&=& \frac{\mu_{bu}-\mu_{rs}}{2}= -\frac{\mu_e}{2}-
\frac{\mu_3}{4}-\frac{\mu_8}{2}+\frac{\mu_S}{2},\nonumber \\
\delta \mu_{III}&=& \frac{\mu_{rd}-\mu_{gu}}{2}= \frac{\mu_e}{2}+
\frac{\mu_3}{2}, \label{dmu}
\end{eqnarray}
respectively, where Eq.(\ref{muk}) has been considered for the
strange quark. It is well known that gapless formation becomes
possible when the mismatch exceeds the gap. By inserting the
solutions Eqs.(\ref{ncfl}) and (\ref{ncflk}) into Eq.(\ref{dmu}), we
will investigate the critical conditions for gapless formations in
CFL and CFL$K^0$, respectively.
%Similarly, the average chemical potential for a given pairing may be
%obtained from $\bar{\mu}_{\alpha i-\beta j}=(\mu_{\alphai}+\mu_{\beta j})/2$.
%Before going to be specifics, let us briefly review the mechanism
%for the gapless formation

In CFL matter, the mismatches for the three channels are
\begin{eqnarray}
\delta \mu_I= \mu_S,\,\,\,\, \delta \mu_{II}= \mu_{S}
-\mu_e,\,\,\,\,\, \delta \mu_{III}= \mu_e,
\label{dmucfl}\end{eqnarray} respectively, while the average
chemical potentials defined by $\bar{\mu}_{\alpha i-\beta
j}=(\mu_{\alpha i}+\mu_{\beta j})/2$ have the same value, i. e.
$\bar{\mu}_{I}=\bar{\mu}_{II}=\bar{\mu}_{III}=\mu- \mu_{S}/3$. When
$\delta \mu_I=\mu_S \geq \Delta_0$ the I-channel gapless formation
occurs, since it precedes the phase transition from CFL to unpaired
quark matter \cite{alfj02,alfd}.  Note that not merely the pairings,
but also the I-channel involved modes ( i. e. $bu$ and $gs$ )
themselves are $\tilde{Q}$ neutral. If only the I-channel gapless
phenomenon exists, thus, the gapless phase remains to be a
$\tilde{Q}$ insulator, as shown in the region between the dot and
dashed lines in Fig. 1.
%If the conditionfor the II-channel gapless formation is not satisfied, the gapless
%with only the I-channel gaplessphenomenon r
% once gapless modes exist in II channel. As
%the , the so-called gCFL phase
%by including .
%As shown in Fig. 1(b), the region above the dashed line remains to
%be the CFL$K^0$ phase while the gCFL$K^0$ phase might exist in the
%region between the dashed and dot lines.
On the other hand, the II-channel gapless formation becomes possible
when $\delta \mu_{II}=\mu_S -\mu_e \geq \Delta_0$. Once the
II-channel modes with nonzero $\tilde{Q}$ charge exist, the
resulting phase is no longer an insulator. In this sense, the
so-called gCFL phase is a $\tilde{Q}$ conductor and the additional
electron density is required to guarantee its $\tilde{Q}$
neutrality. As stressed in Ref.\cite{alf04}, the gCFL location in
the $(\mu_S,\mu_e)$ plane is limited on a single line that very
close to the dashed line of Fig. 1. Because the dashed line comes
from the critical condition $\mu_S=\mu_e +\Delta_0$, the gapless
strength in II channel is almost infinitesimal while the I-channel
gapless phenomenon actually dominates the physics of gCFL.
%%%%% gap variation in gcfl
%In this case, thereexist gapless ( unpaired ) modes in this channelso that the
%corresponding gap $\Delta_1$ becomes smaller than the original value$\Delta_0$.
%the gap $\Delta_2$ is almost unchanged with respect to$\Delta_0$ and
%%%%%%%%%%%%%%%%

Now we consider the mismatches in the CFL$K^0$ environment to
demonstrate how the modified neutral solution Eq.(\ref{ncflk})
works. Different from Eq.(\ref{dmucfl}), the mismatches for the
three channels are
\begin{eqnarray} \delta \mu_I=
\frac{\mu_S}{2},\,\,\,\, \delta \mu_{II}= \frac{3\mu_S}{4}
-\mu_e,\,\,\,\,\, \delta \mu_{III}= \mu_e-\frac{\mu_S}{4},
\label{dmucflk}\end{eqnarray} respectively, while the average
chemical potentials are
%no longer equal to each other and they are
\begin{eqnarray}
\bar{\mu}_{I}=\mu- \frac{\mu_S}{3},\,\,\,\, \bar{\mu}_{II}=\mu-
\frac{7\mu_S}{12},\,\,\,\, \bar{\mu}_{III}=\mu- \frac{\mu_S}{12}.
\label{amucflk}\end{eqnarray} According to Eq.(\ref{dmucflk}), the
critical condition for the I-channel gapless formation is
$\mu_S=2\Delta_0$. When $\mu_S$ is close to $2\Delta_0$, however,
the phase transition to unpaired quark matter occurs and the
I-channel gapless phenomenon no longer appears accordingly.
\footnotemark[1] \footnotetext[1] { Strictly speaking, the phase
transition to unpaired matter is slightly delayed in the presence of
$K^0$ condensation. By taking into account the  $K^0$ condensed free
energy as well as the changes in the common Fermi momenta
Eq.(\ref{amucflk}), we have reexamined this transition in the
similar way as Ref.\cite{alfj02}. To order $m_s^4$, it is found that
the critical value of $\mu_S$ is $2\Delta_0/\sqrt{1-2
f_\pi^2/3\mu^2}\approx 2\Delta_0$, where the in-CFL-medium decay
constant $f_\pi$ is about $0.2\mu$ \cite{fpi}.}

On the other hand, the II-channel gapless formation is still
possible, but the critical condition is modified as
\begin{eqnarray}  \mu_S = \frac{4\Delta_0}{3}+\frac{4\mu_e}{3}.
\label{cri}
\end{eqnarray}
If substituting $\mu_e=0$, Eq.(\ref{cri}) means that there exist
gapless modes when $\mu_S \geq 4\Delta_0/3$. In comparison with
$\mu_S \geq \Delta_0$ in the gCFL case, the gapless formation in the
$K^0$ condensed environment is delayed by a factor 4/3, which is
just the conclusion drawn in Ref.\cite{kry}.
%%%%%%%
%This coincidence isnatural since the non-trivial vacuum derived from kaon condensation
%has been considered in \cite{kry}.
%by using the effective Lagrangian forfermions
% and qualitativelyconsistent with that by NJL model \cite{forbes}.
%recently althoughour method is different from that used there.
%. Within an effective theory for fermions, they calculated the spectrum of the
%quasi quark excitations about the kaon condensed state in CFL matter
%and suggested that $m_s^2/2\mu \simeq 4\Delta_0/3$ for the gCFLformation.
%More importantly,  with the II-channel
%gapless phenomenon behaves as a $\tilde{Q}$ conductor.
%%%%%%%\cite{kry,forbes,buba}
Keeping in mind that the corresponding gapless modes are $\tilde{Q}$
charged, the $\tilde{Q}$ neutrality in the resulting gCFL$K^0$ phase
needs to be realized by taking electrons into account explicitly.
Therefore, effects of nonzero $\mu_e$ and electron density on the
gCFL$K^0$ phase must be studied seriously. This issue was not
mentioned in the above-cited literature and might be relevant for
understanding the gapless phase.

%\vspace{0.1cm}\noindent {\bf C.  Electric, color and $\tilde{Q}$
%neutrality in gCFL$K^0$ } \vspace{0.1cm}

For the gCFL$K^0$ phase, the electric/color neutrality condition
Eq.(\ref{neu}) needs to be considered if it exists as a bulk matter.
Note that the total free energy consists of $ \Omega_{CFLK^0}$,
$\Omega_{g}$ and $\Omega_{e}$, where $\Omega_{g}$ denotes the
contribution from gapless modes ( see Sec. III ) and
$\Omega_{e}=-\mu_e^4/12\pi^2$ from electrons. Without losing
generality, we suppose that the values of $\mu_3$ and $\mu_8$ are
deviated from the CFL$K^0$ neutral solution Eq.(\ref{ncflk}) once
gapless phenomenon has occurred. In this case $\Omega_{CFLK^0}$, as
one part of the gCFL$K^0$ free energy, no longer satisfies
Eq.(\ref{neu}) and its derivatives Eqs.(\ref{neueq}) and
(\ref{neueq'}) are usually nonzero. By considering Eqs.(\ref{neueq})
and (\ref{neueq'}), the model-independent results of the
electric/color neutrality in gCFL$K^0$ read
\begin{eqnarray}
%\frac{\partial \Omega_{gCFLK^0}}{\partial\mu_e}
\frac{\mu^2}{12\pi^2}(2\mu_S+3\mu_3+2\mu_8-4\mu_e)+\frac{\partial
\Omega_{g}(\mu_e,\mu_3,\mu_8)}{\partial
\mu_e}-\frac{\mu^3_e}{3\pi^2}=0, \label{neueq2e}
\end{eqnarray}
\begin{eqnarray}
%\frac{\partial \Omega_{gCFLK^0}}{\partial\mu_3}&=&
\frac{\mu^2}{8\pi^2}(2\mu_e-2\mu_3-\mu_S)+\frac{\partial
\Omega_{g}(\mu_e,\mu_3,\mu_8)}{\partial \mu_3}=0,\label{neueq23}
\end{eqnarray}
\begin{eqnarray}
%\frac{\partial \Omega_{gCFLK^0}}{\partial\mu_8}&=&
\frac{\mu^2}{12\pi^2} (2\mu_e-4\mu_8-\mu_S)+\frac{\partial
\Omega_{g}(\mu_e,\mu_3,\mu_8)}{\partial \mu_8}=0,\label{neueq28}
\end{eqnarray}
which correspond to the derivatives of the gCFL$K^0$ free energy
with respect to $\mu_e$, $\mu_3$ and $\mu_8$ respectively.
Obviously, the gCFL$K^0$ neutral solution is deviated from that of
CFL$K^0$ unless $\partial \Omega_{g}/\partial \mu_3$ and $\partial
\Omega_{g}/\partial \mu_8$ are zero.

Besides electric/color neutrality, gCFL$K^0$ is required to be
$\tilde{Q}$ neutral. Note that the free energy $\Omega_{CFLK^0}$ is
independent of $\tilde{Q}$ charge and the $\tilde{Q}$ charge derived
from the gapless phenomenon should be canceled by that from
electrons. By using Eq.(\ref{Q}), the neutrality condition may be
expressed by
\begin{eqnarray} \frac{\partial \Omega_{g}(\mu_e,\mu_3,\mu_8)}{\partial
\mu_e}+\frac{\partial \Omega_{g}(\mu_e,\mu_3,\mu_8)}{\partial
\mu_3}+2\frac{\partial \Omega_{g}(\mu_e,\mu_3,\mu_8)}{\partial
\mu_8}=\frac{\mu_e^3}{3\pi^2},\label{neueqQ}
\end{eqnarray}
where the left-hand side comes from the complete-derivative of
$\Omega_{g}$ with respect to $\mu_{\tilde{Q}}$. Combining
Eqs.(\ref{neueq2e}), (\ref{neueq23}) and (\ref{neueq28}) with
(\ref{neueqQ}), it is found that the derivative
$\partial\Omega_{g}/\partial\mu_8$ is equal to zero so that
$\mu_8=\mu_e/2-\mu_S/4$ actually holds unchanged. Physically, the
reason lies in the facts that only the II-channel gapless phenomenon
occurs in our concerned phase and the combination $Q+2T_8/\sqrt{3}$
becomes zero for the unpaired $ub$ and $rs$ quarks. As a result, the
$\tilde{Q}$ neutrality condition Eq.(\ref{neueqQ}) can be simplified
as
\begin{eqnarray} \frac{\partial
\Omega_{g}(\mu_e,\mu_3,\mu_8)}{\partial \mu_e}+\frac{\partial
\Omega_{g}(\mu_e,\mu_3,\mu_8)}{\partial
\mu_3}=\frac{\mu_e^3}{3\pi^2}.\label{neueqQ2}
\end{eqnarray}
%%%%%%%%%%
%It is worthy being stressed that both electron and color chemical
%potentials are associated with $\mu_{\tilde Q}$ ( see ). Now that
%$\Omega_{g}$ and $\Omega_{e}$ contribute nonzero $\tilde{Q}$charge,
%Essentially, the $\tilde{Q}$ neutral condition should be consistent
%with the electric/color neutral condition.
%Eq.(\ref{neueqQ}) Eqs.(\ref{neueq2e}), (\ref{neueq23}) and (\ref{neueq28}).

Further inserting Eq.(\ref{neueqQ2}) into (\ref{neueq2e}), we find
that no more result than Eq.(\ref{neueq23}) can be obtained. It
means that the relation $\mu_3=\mu_e-\mu_S/2$ may change as the
gapless phenomenon occurs. In the gCFL$K^0$ phase, the value of
$\mu_3$ must be solved from Eq.(\ref{neueq23}) ( and
Eq.(\ref{neueq2e}) equivalently ) numerically. It is worthy being
stressed that the above results are still obtained in a model
independent way and hold strictly valid at leading order. In a
model-dependent treatment, color chemical potentials are usually
given " by hand " to guarantee electric/color neutrality ( see e. g.
Ref.\cite{alf04} ). To examine the II-channel gapless phenomenon
self-consistently, we take the possible change of $\mu_3$ into
account and thus adjust the relative and average chemical potentials
to be
\begin{eqnarray} \delta \mu= -\frac{3\mu_e}{4}- \frac{\mu_3}{4}
+\frac{5\mu_S}{8},\label{du}
\end{eqnarray}
and
\begin{eqnarray}
\bar{\mu} = \mu- \frac{\mu_e}{4}+ \frac{\mu_3}{4}
-\frac{11\mu_S}{24}, \label{au}
\end{eqnarray}
for II channel, which make sense only if $\delta \mu_{II}\geq
\Delta_0$ is satisfied; otherwise, no gapless formation occurs and
$\mu_3$ holds unchanged.

\vspace{0.2cm}\noindent {\bf III. PHASE TRANSITION TO THE gCFL$K^0$
PHASE }\vspace{0.2cm}

%\vspace{0.1cm}\noindent {\bf A. Phase transition to the gCFL$K^0$
%phaseGAPLESS FORMATION IN THE $K^0$ CONDENSEDENVIRONMENT } \vspace{0.1cm}

As pointed out in Refs.\cite{alf04,huang}, the gapless phase
corresponds to the common solution of electric/color neutrality and
gap equation. By using Eqs.(\ref{du}) and (\ref{au}), let 's
consider the gap variation and the gap equation in the gCFL$K^0$
phase. When gapless modes appear, the gaps for various pairings
separate from each other and their values should be solved from
three gap equations, namely ${\partial \Omega}/{\partial
\Delta_1}={\partial \Omega}/{\partial \Delta_2}={\partial
\Omega}/{\partial \Delta_3}=0$, in principle. In view of the fact
that the II-channel gapless phenomenon dominates the physics of
gCFL$K^0$, the variation of $\Delta_2$ from the original value
$\Delta_0$ is obvious while the changes in $\Delta_1$ and $\Delta_3$
are small relatively. Now we ignore the latter variations as well as
the $\mu_S$ dependence of $\Delta_0$ so as to simplify three gap
equations into one equation.
%can be treated in a model-independent way.

For the II-channel pairing, the dispersion relation of quasiquarks
takes the form of
\begin{eqnarray} E(p)= |\delta\mu \pm
\sqrt{(p-\bar{\mu})^2+\Delta_2^2}|.\label{ep}
\end{eqnarray}
It is easily found that the excitation energy for quasiquarks
becomes zero at the momenta
\begin{eqnarray} p^{\pm}= \bar{\mu}\pm
\sqrt{{\delta\mu}^2-\Delta_2^2}.\label{pm}
\end{eqnarray}
When $\delta\mu \geq \Delta_2$ Eq.(\ref{pm}) makes sense and there
exist gapless modes in the blocking region $p\in(p^-,p^+)$. As a
consequence, the gapless modes provide
\begin{eqnarray} {\Omega}_{g}=
- \int_{p^{-}}^{p^{+}} \frac{p^2 dp}{2\pi^2}
[\delta\mu-\sqrt{(p-\bar{\mu})^2+\Delta_2^2}], \label{pgap}
\end{eqnarray}
to the total free energy for gCFL$K^0$. Besides the contribution
${\Omega}_{g}$, the free energy from the CFL pairings is influenced
by the gap variation. In the CFL matter with the common gaps, it is
well known that such a free energy is equal to $- {3\Delta_0^2
\mu^2}/{\pi^2}$ at the leading order \cite{alfj02}. Since only the
binding energy from the II-channel pairing is influenced within our
assumption, the gap variation leads to the additional contribution
\begin{eqnarray} \frac{(\Delta_0^2- \Delta_2^2) \bar{\mu}^2}{\pi^2},\label{ppair}
\end{eqnarray}
to the pairing free energy. Together with Eqs.(\ref{pgap}) and
(\ref{ppair}), the gap equation can be expressed by
\begin{eqnarray}
%\frac{\partial \Omega_{gCFLK^0}}{\partial\Delta_2}=
-\frac{2\Delta_2 \bar{\mu}^2}{\pi^2}+\frac{\partial
\Omega_{g}}{\partial \Delta_2}=0,\label{geq}
\end{eqnarray}
for the given values of $\mu_e$, $\mu_3$ and $\mu_8$. As long as the
II-channel gapless strength is not-very-large, the simplified gap
equation is valid for reflecting the gap variation in gCFL$K^0$.
Also, it is compatible with the gCFL$K^0$ neutrality condition since
Eqs.(\ref{du}) and (\ref{au}) have been considered in the free
energy $\Omega_g$.

With the help of the  gCFL$K^0$ neutral solution and the gap
equation, we turn to examine the location of gCFL$K^0$ in the
$(\mu_S,\mu_e)$ plane. Different from the gCFL case, there is no
reason to anticipate that the gCFL$K^0$ location is limited in the
vicinity of the critical line obtained from Eq.(\ref{cri}) ( the
dashed line in Fig. 2 ). In fact, if gCFL$K^0$ located near this
line, the free energy $\Omega_{g}$ would be almost infinitesimal (
remember only the II-channel gapless phenomenon exists ). In that
case, the resulting gapless phase might be hard to become
energetically favorable with respect to the CFL$K^0$ phase.
%Therefore, Eq.(\ref{cri}) is no longer the sufficient condition for
%the gCFL$K^0$ formation.
As the matter satisfying the neutrality and
gap equation, gCFL$K^0$ becomes actually possible as long as it has
a lower free energy ( a higher pressure ) than CFL$K^0$. Thus, the
CFL$K^0$-gCFL$K^0$ phase transition line is obtained from the Gibbs
condition of pressure equilibrium, i. e.
\begin{eqnarray} \delta {\cal P}= -\frac{(\Delta_0^2-\Delta_2^2) \mu^2}{\pi^2}-
\Omega_{g}+\frac{\mu_e^4}{12\pi^2}=0.\label{gibbs}
\end{eqnarray}
%for given $\mu_S$ and $\mu_e$.
As shown in Fig. 2, the critical line for the gapless formation (
the dashed line ) and the phase transition line ( solid ) separate
from each other  obviously. When $m_s^2/2\mu$ is large relatively,
the gCFL$K^0$ phase with $\delta {\cal P} \geq 0$ is found to exist
in the region below the solid line. Correspondingly, the system
remains to be CFL$K^0$ in the region between the solid and dashed
lines.

The Gibbs condition Eq.(\ref{gibbs}) implies that the phase
transition is of first-order. \footnotemark[2] \footnotetext[2] {
%%%%%
%Strictly speaking, it is a single sharp interface between the
%$\tilde{Q}$-free CFL$K^0$ and $\tilde{Q}$-dependent gCFL$K^0$
%matter, similar to that happens between insulator and metal. In
%%sense of an insulator-conductor interface.
%  As the phase boundary, the solid line separates
%Note thatCFL$K^0$ is the insulator in the sense of electric and $\tilde{Q}$
%charges, while strongly dependent of theexistence of $\mu_e$.
%%%%%%%
The physical picture is valid at leading order. Since $m_s^2/2\mu$
has been treated as an external chemical potential, there do exist
not only one chemical potential in our concerned system. Thus it is
natural that the transition to the gapless phase is first order
while the gapless formation itself is continuous ( see e. g.
Ref.\cite{glen} for an analogy ). } This is very different from the
previous conclusion that the CFL-gCFL phase transition is of
second-order ( at zero temperature ) \cite{alf04}. In fact, the gCFL
phase with the I- and II-channel gapless phenomena is \emph{barely}
a conductor while gCFL$K^0$ with only the I-channel one is a
conductor at all. In the gCFL$K^0$ case, no extra $\tilde{Q}$
neutral condition ( such as $\mu_e^3 \sim (p^+-p^-){\bar{\mu}}^2$ in
the gCFL case \cite{alf04} ) is required.
%Even if $\mu_e$ is not-very-large, thus, the gapless strength is no
%longer enforced to be infinitesimal.
As a consequence,  the $\tilde{Q}$-neutral gCFL$K^0$ matter appears
in a finite region of the $(\mu_S,\mu_e)$ plane rather than just at
a single line. Also, the quadratic dispersion relation predicted in
gCFL \cite {alf04} no longer valid and the gapless modes in
gCFL$K^0$ might have a linear dispersion relation as the
conventional gapless phase.
%%%
%When $m_s^2/2\mu$ is small relatively, the gapless strength is so
%tiny that the $\tilde{Q}$ charge derived from gCFL$K^0$ is almost
%neglectable. In this case, the electron chemical potentials are zero
%approximately, as shown by the solid line in Fig. 2. Once
%$m_s^2/2\mu$ becomes large, the electron chemical potentials have
%finite value and the nature of a first-order phase transition
%manifests. As a consequence,
%%%%%%%%%%%%%%%%%%%%%%%%%%
%%%%%%%%%%%%%
%\vspace{0.1cm}\noindent {\bf B. Chromomagnetic instability
%reexamined} \vspace{0.1cm}
%the novel phase structure of gCFL$K^0$ might be helpful for
%overcoming chromomagnetic instability. Instead of calculating the
%Debye and Meissner masses for gluons,

Finally, we analyze the possibility of eliminating the
chromomagnetic instability happens in gCFL qualitatively. In the
gCFL case, the I- and II-channel gapless formations coincide at the
point of $(\mu_e,\mu_S)=(0,\Delta_0)$, i. e. the $gs$-$bd$ and
$rs$-$bu$ pairings become breached simultaneously. As stressed in
Refs.\cite{cas05,fuku}, this feature is responsible for the
instability occurs for $A_1$ and $A_2$ gluons : since the
$m_s$-relevant self energy for $A_{1,2}$ stems from the loop diagram
composed of $gs$ and $rs$ quarks, the possible coexistence of
gapless ( unpaired ) $gs$ and $rs$ modes provides a singular
contribution to the $A_{1,2}$ Meissner masses. In the gCFL$K^0$
case, however, the I-channel gapless formation does not occur yet.
Therefore, there should exist no longer singularities in the
self-energy for $A_{1,2}$ Meissner masses and the corresponding
instability should disappear in the gCFL$K^0$ phase.
%%%%%%%%
%Technically
%$A_1$ and $A_2$ should do not exhibit chromomagnetic
% also. Now that there is not otherinstabilities
%, say these . In this sense, we conclude
On the other hand, the feature that the gCFL location is very close
to its critical line has been pointed out to be responsible for the
instability for $A_3$, $A_8$ gluons and $A_\gamma$ photon  mainly
\cite{fuku}. For instance, the ( almost ) infinitesimal II-channel
gapless strength in gCFL leads to the singularities in the
$A_{3,8,\gamma}$ masses \cite{fuku}. In the gCFL$K^0$ case, however,
the location is not limited in the vicinity of the critical line.
Instead, a finite region in the $(\mu_S,\mu_e)$ plane is available
for the gCFL$K^0$ existence. Thus such kind of instability might no
longer appear at least for somewhat region in the $(\mu_e,\mu_S)$
plane.

Since only $A_{1,2}$ and $A_{3,8,\gamma}$ exhibit imaginary Meissner
masses in gCFL, the previous-predicted instability is very likely to
be removed in the gCFL$K^0$ phase. Physically, it can be understood
from the viewpoint that chromomagnetic instability is perhaps
%likely to originate from theimproper-established vacuum
eliminated by including the proper vacuum. It has been suggested
that nonzero vacuum expectation values of gluon such as $\langle
A^0_3\rangle$ and/or $\langle A^0_8\rangle$ are helpful for removing
the gCFL instability \cite{cas05}. In Ref.\cite{gorb}, the
instability in two-flavor superconductor was argued to be resolved
by gluon condensation.
%More recently,Sch\"{a}fer has stressed the stability of the kaon condensed phase
%against the formation of gapless fermions by using an effective
%theory \cite{sch06}.
In the present work, the nonzero $\langle A^0_3\rangle$ and $\langle
A^0_8\rangle$ ( anisotropic vacuum ) derived from the kaon
condensation have been attributed to the change of the
electric/color neutral solution and then been considered in the
study of the gapless formation. In this sense, the conjecture that
gCFL$K^0$ is out of chromomagnetic instability partly is not very
surprising. However, intrinsic links between the ( $p$-wave ) kaon
supercurrent phase establised in \cite{sch06} and the
present-discussed gCFL$K^0$ phase still remain to be clarified.
%which is left in further work.
% unclear whether chromomagneticinstability is a general property of the phases with gapless modes.
%Even if the gCFL singularities have been eliminated
Also, we could rule out the possibility that instabilities
especially these occur for $A_4$, $A_5$, $A_6$ and $A_7$ gluons
arise again. The quantitative calculation of the gluonic Meissner
masses in gCFL$K^0$ is still necessary which is beyond the scope of
the present work.

%\vspace{0.2cm}\noindent {\bf IV. SUMMARY AND OPEN QUESTIONS
%\vspace{0.2cm}\noindent {\bf IV. SUMMARY }\vspace{0.2cm}

In summary, we investigated electric/color neutrality and gapless
formation in the CFL matter with $K^0$ condensation. By taking the
CFL$K^0$ neutral solution into account, we clarify why the gCFL$K^0$
formation is delayed in comparison with the gCFL case. More
importantly, it is found that
%some novel features is found in theresulting gapless phase. Firstly,
the gapless phenomenon for down-strange pairing ( I channel ) is
absent while that for up-strange one ( II channel ) becomes
dominated. After the model-independent treatments of the gCFL$K^0$
neutrality and the gap equation, we suggest that the
CFL$K^0$-gCFL$K^0$ phase transition is first order. The novel phase
structure implies that the previous-predicted gCFL instability might
be removed at least partly. These conclusions are likely to be
important for fully understanding the unconventional CFL phases in
the presence of $m_s$ and $\mu_e$. Even if the variations of three
gaps are considered in a model-dependent calculation, the current
results should be qualitatively correct. Of course, there are still
some unanswered aspects in the present work. First of all, the $K^0$
condensed CFL matter has been treated as the background of gapless
formation, which is safe only if $\mu$ is not too small ( otherwise,
the condensation can easily supressed by instanton effect \cite{ins}
). Due to the nature of kaon condensation, nevertheless, the
gCFL$K^0$ phase with only II-channel gapless phenomenon is difficult
to gain an obviously lower free energy than the gCFL phase.
% Also, the former is likely to be \emph{not} energetically
%favorable with respect to the latter.  If so
In this sense, our suggested gapless phase
%the reason should lie in the fact that only the
%II-channel gapless phenomenon exists in gCFL$K^0$ and the
%corresponding gapless strength is small relatively, also. In this
%sense, the present-predicted gapless phase
could not replace the role of gCFL, in particular its role in
neutron star cores \cite{alf05}, if ignoring the gCFL instability.
Secondly, more physics involving Goldstone-mode condensation is not
discussed. In fact, we could not exclude possibilities of the
charged kaon ( and even other modes ) condensations completely. For
instance, if $\mu_e$ is large enough and the $K^-$ condensation
occurs, the gapless phenomenon for up-down pairing ( the III channel
) needs to be included seriously.
%%%%
%Another important issue that has not discussed in this work involves
%the role of In addition, the phase transitions from the gCFL$K^0$ to
%non-CFL phases need to be examined. The transition to the unpaired
%quark matter is expected to be still occurs in the vicinity of
%$\mu_S=2\Delta_0$, while the transition(s) to various two-flavor
%quark superconductors may be more complicated and warrants much
%further investigation.
%Finally, it is noted that the present results are of the leading
%order. Further investigations should treat effects of the strange
%quark mass and the gap equations more seriously as well as take the
%symmetric color-sextet pairing into account, since these have been
%considered in the NJL model calculations of kaon condensation
%\cite{forbes,buba}.
Some of the above-mentioned problems are being investigated.
%%%%%%%%
% where $m_s^2/2\mu \equiv \mu_S$ is treated as a small,external chemical potential.
% the density dependence of the color
%superconducting gap and construct the dynamical quark mass within a
%more realistic framework such as that beyond the bag model.
%With relatively large $\mu_S$,
%the changes of the strange content in either CFL$K^0$ or gCFL$K^0$
%might be too obvious to be ignored.
%and the gapless formation obtained in thepresent work are influenced
% sincethe kaon mass is related with the CFL gaps. In this case, the
%leading-order results obtained in this work might be not strictly
%correct also.
%%%%%%%%%%%
% Alternatively,the possibility of the mixed phase consisting of the gapless
%color-flavor locked and unpaired matter needs to be reexamined since
%gCFL$K^0$ is a conductor at all. As for the phase
%( although 2SC was doubted to be chromoinstablealso \cite{instab} ).
%Works along these directions are worth being pursued.

\vspace{0.5cm} \noindent {\bf Acknowledgements} \vspace{0.5cm}

This work was supported by National Natural Science Foundation of
China ( NSFC ) grant 10405012 and DOE grant DE-FG02-87ER40328.

\newpage
\begin{figure}
%\begin{center}
%\includegraphics*{gkf1.eps}
\caption{ Schematic phase structure of the gCFL formation in the
$(\mu_e, \mu_S \equiv m_s^2/2\mu)$ plane, where the critical
conditions for I- and II-channel gapless phenomena and the phase
transition to unpaired quark matter are shown by the dot, dashed and
dot-dashed line respectively. The $\tilde{Q}$-neutral gCFL phase
locates below but very close to the dashed line \cite{alf04}.}
% andthe III-channel gapless phenomenon is ignored.}
%\end{center}
\end{figure}

\begin{figure}
%\begin{center}
%\includegraphics*{gkf2.eps}
\caption{ Similar as Fig.1, but for the gCFL$K^0$ formation. The
I-channel critical line ( dot ) coincides with the line of the phase
transition to unpaired matter approximately, while the solid line
corresponds to the CFL$K^0$-gCFL$K^0$ phase transition ( see text
).}
%\end{center}
\end{figure}


\vspace{0.7cm}

\vspace{0.2cm}

\begin{thebibliography}{99}
\bibitem{alf}
 M. Alford, K. Rajagopal and F. Wilczek, Phys. Lett. {\bf B422}, 247 (1998);
 M. Alford, K. Rajagopal and F. Wilczek, Nucl. Phys. {\bf B537}, 443 (1999).
\bibitem{alf03}
For reviews, see K. Rajagopal and F. Wilczek, hep-ph/0011333; T.
Sch\"{a}fer, hep-ph/0304281; M. Alford, Prog. Thero. Phys. Suppl.
{\bf 153}, 1 (2004); D. H. Rischke, Prog. Part. Nucl. Phys. {\bf
52}, 197 (2004).
\bibitem{liu}   E. Gubankova, W. V. Liu and F. Wilczek, Phys. Rev. Lett. {\bf
91}, 032001 (2003)
\bibitem{alf04}
 M. Alford, C. Kouvaris and K. Rajagopal, Phys. Rev. Lett. {\bf 92}, 222001
 (2004); see also,  M. Alford, C. Kouvaris and K. Rajagopal, Phys. Rev. D
{\bf 71}, 054009 (2005).
\bibitem{huang} I. Shovkovy and M. Huang, Phys. Lett. {\bf B564},
205 (2003).
\bibitem{cas05}
R. Casalbuoni, D. Gatto, M. Mannareli, G. Nardulli and M. Ruggieri,
Phys. Lett. {\bf B605}, 362 (2005); \emph{ibid} {\bf B615}, 297
(2005).
\bibitem{fuku}
K. Fukushima, Phys. Rev. D {\bf 72}, 074002 (2005).
\bibitem{alfjp05}
M. Alford and Q. H. Wang, J. Phys. G {\bf 31}, 719 (2005).
\bibitem{sch}
P. F. Bedaque and T. Sch\"{a}fer, Nucl. Phys. {\bf A697}, 802
(2002).
%; see also,T. Sch\"{a}fer, Phys. Rev. Lett. {\bf 85}, 5531 (2000).
%Sch\"{a}fer, Nucl. Phys. {\bf A702}, 167c (2002).T. Sch\"{a}fer, Nucl. Phys. {\bf A575}, 269 (2000);
\bibitem{kr}
D. B. Kaplan and S. Reddy, Phys. Rev. D {\bf 65}, 054042 (2002); S.
Reddy, M. Sadzikowski and M. Tachibana, Phys. Rev. D {\bf 68},
053010 (2003).
\bibitem{kry}
A. Kryjevski and T. Sch\"{a}fer, Phys. Lett. B {\bf 606}, 52 (2005);
see also, A. Kryjevski and D. Yamada, Phys. Rev. D {\bf 71}, 014011
(2005).
\bibitem{forbes} M. M. Forbes, Phys. Rev. D {\bf 72}, 094032 (2005).
\bibitem{buba} M. Buballa, Phys. Lett. B {\bf 609}, 57 (2005).
\bibitem{sch06}
T. Sch\"{a}fer, Phys. Rev. Lett {\bf 96}, 012305 (2006).
\bibitem{alfj02}
 M. Alford and K. Rajagopal, J. High Energy Phys. {\bf 0206}, 031 (2002);
\bibitem{rw01}
 K. Rajagopal and F. Wilczek, Phys. Rev. Lett {\bf 86}, 3492 (2001).
\bibitem{kry03} A. Kryjevski, Phys. Rev. D {\bf 68}, 074008 (2003).
\bibitem{alfd}
M. Alford, K. Rajagopal, S. Reddy and F. Wilczek, Phys. Rev. D {\bf
64}, 074017 (2001).
\bibitem{fpi}
D. T. Son and M. Stephanov, Phys. Rev. D {\bf 61}, 074012 (2000);
\emph{ibid}{\bf 62}, 059902 (2000).
\bibitem{glen} N. K. Glendenning, Phys. Rev. D {\bf 46}, 1274 (1992).
\bibitem{gorb}
E. V. Gorbar, M. Hashimoto and V. A. Miransky, Phys. Lett. B {\bf
632}, 305 (2006)
\bibitem{ins}
T. Sch\"{a}fer, Phys. Rev. D {\bf 65}, 094033 (2002).
\bibitem{alf05}
M. Alford, P. Jotwani, C. Kouvaris, J. Kundu and K. Rajagopal, Phys.
Rev. D {\bf 71}, 114011 (2005).
%\bibitem{instab}M. Huang and I. Shovkovy, Phys. Rev. D {\bf 70},
%051501(R) (2004); Phys. Rev. D {\bf 70}, 094030 (2004).
%\bibitem{zhang}X. B. Zhang and X. Q. Li, Phys. Rev. D {\bf 72}, 054021 (2005).
%\bibitem{zhang}X. B. Zhang, Y. Luo and X. Q. Li, Phys. Rev. D {\bf 68}, 054015(2003).
%\bibitem{zhang2}X. B. Zhang and X. Q. Li, Phys. Rev. D {\bf 70}, 054010 (2004).
\end{thebibliography}
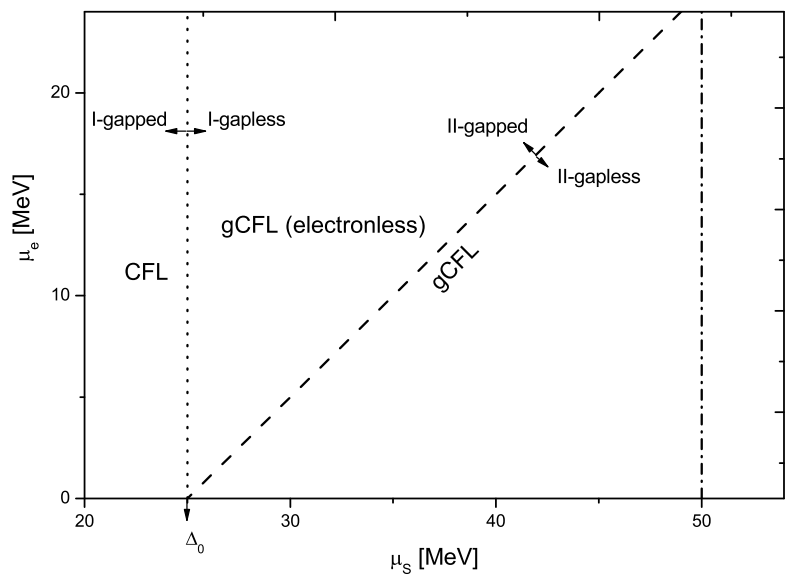
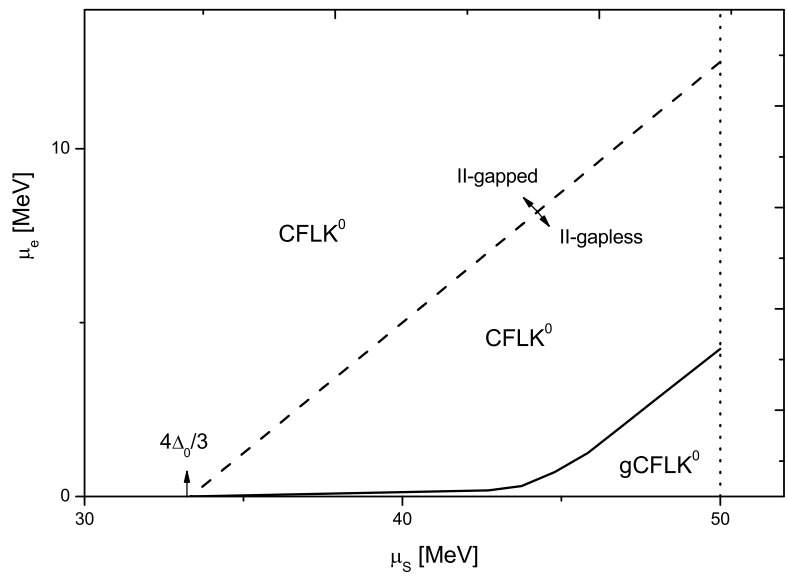
\end{document}